\documentstyle[twoside,12pt,epsf]{article}
\normalsize
\newcommand{\bdi}{\begin{displaymath}}
\newcommand{\edi}{\end{displaymath}}
\newcommand{\bfi}{\begin{figure}}
\newcommand{\efi}{\end{figure}}

\newcommand{\beq}{\begin{equation}}
\newcommand{\eeq}{\end{equation}}
\newcommand{\beqa}{\begin{eqnarray}}
\newcommand{\eeqa}{\end{eqnarray}}

\newcommand{\wt}{\widetilde}
\newcommand{\wh}{\widehat}
\newcommand{\tr}{{\rm tr}\, }

\begin{document}
\title{Covariant Schwinger Terms }
\author{{\bf C. Adam, C. Ekstrand, T. S\'ykora}}
\date{}
\maketitle
\newcommand{\Ref}[1]{(\ref{#1})}
\newcommand{\F}{{\cal F}}
\newcommand{\A}{{\cal A}}
\newcommand{\G}{{\cal G}}
\begin{abstract}
There exist two versions of the covariant Schwinger term in the 
literature. They only differ by a sign. However, we shall show 
that this is an essential difference. We shall carefully 
(taking all signs into account) review the existing 
quantum field theoretical computations for the covariant 
Schwinger term in order to determine the correct expression.
\end{abstract}


\section{Introduction}

One essential feature of chiral gauge theories is the violation of gauge
invariance when chiral (Weyl) fermions are quantized. This loss of
gauge invariance results in a non-invariance of the vacuum functional in
an external gauge field under gauge transformations
and in a non-conservation (anomalous divergence)
of the corresponding gauge current in a Lagrangian (or space-time) 
formulation
(``anomaly'' \cite{Ad1}--\cite{Ba1}), 
or in anomalous contributions to the equal-time 
commutators of the
generators of time-independent gauge transformations in a Hamiltonian
formulation (``Schwinger term'' or ``commutator anomaly'', 
\cite{JJ1}--\cite{FaSh2}).
One regularization scheme, where all the anomalous terms are related to
functional derivatives of the vacuum functional, leads to the so-called
consistent anomalies. These anomalies have to obey certain consistency 
conditions because of their relation to functional derivatives
\cite{WZ1}. One way
of determining these consistent anomalies (up to an overall constant) is
provided by the descent equations of Stora and Zumino \cite{Sto1,Zu1}.
They provide a simple algebraic scheme -- based on some geometrical
considerations -- for the computation of the consistent chiral anomaly
in the space-time formalism and for the corresponding equal-time 
commutator anomaly, the Schwinger term (as well as for higher cochain
terms). 

On the other hand, it is possible to choose a gauge-covariant
regularization for the gauge current. This covariant current cannot be
related to a functional derivative of the vacuum functional (because of
the gauge non-invariance of the latter). As a consequence, the covariant
current anomaly does not obey the
consistency condition. Nevertheless, there exists a covariant
counterpart for each consistent cochain in the descent equations.
The first derivation of an algebraic computational scheme for covariant 
cochains appears to be the one by Tsutsui \cite{Tsu1}, using the anti-BRST
formalism. Further, a covariant version of the descent equations was
derived by Kelnhofer in \cite{Keln1}. The covariant cochains resulting
from the calculations by Tsutsui on one hand, and by Kelnhofer on the
other hand are, in fact, different, as was shown in \cite{Ek1}. 
Tsutsui's and Kelnhofer's formulas predict the same anomaly in
space-time, but their Schwinger terms differ by a sign. 
The higher cochains (with more than 2 ghosts) seem to be unrelated. 
We shall give an answer to which of the two formulas is correct, in the
sense that it is reproduced by a full quantum field theoretic
calculation. 

The easiest way to do this calculation would be to compute one of the
higher covariant anomalies in some quantum field theoretic setting,
because these higher covariant anomalies are given by completely
different expressions in \cite{Tsu1} and \cite{Keln1}. However, 
although it has been claimed that the higher cochains can have a 
physical meaning, this is far from understood. 
It is therefore not sound to use these terms to argue 
which of the two formulas is correct. Instead, we shall use the  
sign of the Schwinger term as a referee. 

We shall use three methods to determine the 
correct expression for the covariant Schwinger term. 
They all have to  
be used with care since we are after the 
sign difference between Tsutsui's and Kelnhofer's 
predictions. 
The first two methods are to apply the quantum field theoretical 
calculation schemes that have been used by
Adam \cite{Adam1} and by Hosono and Seo \cite{HS}, respectively.

The third method is the one by Wess \cite{Wess1},  
relating the Schwinger term (consistent or covariant) 
in any even dimensional space-time with the corresponding 
space-time anomaly.  This method was
used by Schwiebert \cite{Schw1} for the consistent case and by Kelnhofer
\cite{Keln2} in the covariant case. Thus, by using the 
expression for the covariant anomaly (which everyone 
agrees on) the covariant Schwinger term can be determined. 
Again, care has to be taken. For this, we first perform 
the calculation in the consistent formalism and set 
conventions so the result agrees with what is predicted by 
the descent equations. The corresponding covariant 
computation is then to determine the covariant Schwinger 
term including the sign. We shall perform these calculation 
in 1+1 and 3+1 dimensions. 

From the explicit calculations we find that our quantum 
field theoretical methods produce the same expression (i.e. sign) as 
Kelnhofer's covariant descent equations. This result is not obvious 
since Kelnhofer's approach, as well as Tsutsui's, seems to be based only
on the requirement of covariance. We shall however show that there is 
in fact a natural interpretation of the Kelnhofer formula, as one would
expect.

Our paper is organized as follows. In Section 2 we briefly describe the
geometrical setting for the description of anomalies and review the
derivation of the consistent and covariant chain terms. In the covariant
case, both Tsutsui's and Kelnhofer's versions of the chain terms are
given and a geometrical description of Kelnhofer's construction is  
provided. In Section 3 the consistent and covariant Schwinger terms in
1+1 dimensions are calculated using the methods of \cite{Adam1} and of
\cite{HS}. Finally, in Section 4 the Schwinger terms in 1+1 as well as
in 3+1 dimensions are calculated with the help of the method of Wess
\cite{Wess1}. 

\section{Consistent and covariant cochains}

We shall start with deriving the consistent 
chiral anomaly for a non-abelian gauge theory. Consider therefore 
Weyl fermions $\psi$ coupled to an external gauge field $A\in\A$. 
$\A$ is the affine space of of gauge connections and the gauge group 
$G$ is assumed to be a compact, semi-simple matrix group. We assume that 
the space-time $M$ is a smooth, compact, oriented, even-dimensional 
and flat Riemannian spin manifold without boundary. The group $\G$ 
of gauge transformations consists of diffeomorphisms of a principal 
bundle $P\stackrel{G}{\rightarrow }M$ such that the base remains 
unchanged. It acts on $\A$ by pull-back and to make this action free 
we restrict to gauge transformations that leaves a reference point 
$p_0\in P$ fixed.

The generating functional is given by 
\begin{equation}
\exp (-W(A))=\int _{\psi ,\bar{\psi }}\exp (-\int _M\bar{\psi }
\partial \!\!\! /_A^+\psi d^{2n}x),
\end{equation}
where $W$ is the effective action and $\partial \!\!\! /_A^+ =
\partial \!\!\! /_A(1+\gamma _5)/2=\gamma ^\mu (\partial _\mu 
+A_\mu )(1+\gamma _5)/2$. We shall use conventions such that 
$\gamma ^\mu $ is hermitean and $A_\mu$ is anti-hermitean. It has 
been argued that a correct interpretation of the generating functional 
is as a section of the determinant line bundle $\mbox{DET}i\partial 
\!\!\! /_A =\mbox{det ker}i\partial \!\!\! /_A^+\otimes 
(\mbox{det coker}i\partial \!\!\! /_A^+)^\ast$. It can be viewed as a 
functional by comparing with some reference section. Associated with the 
determinant line bundle is a connection with corresponding curvature
\begin{equation}
\label{eq:CU}
F=-2\pi i \frac{1}{(n+1)!}\left(\frac{i}{2\pi }\right) ^{n+1}
\int _M\mbox{tr}\left( \F ^{n+1}\right) , 
\end{equation}
\cite{BF}. 
Here, $\F =(d+\delta)(A+(d_A^\ast d_A)^{-1}d_A^\ast )+
(A+(d_A^\ast d_A)^{-1}d_A^\ast )^2$ is a curvature of the principal 
bundle $P\times\A\rightarrow M\times \A$ and $\delta$ is the exterior 
differential in $\A$. The choice of $\F $ is motivated by gauge 
invariance of the determinant line bundle, \cite{AS,Ek3}.

Recall that 
\begin{equation}
\label{eq:FO}
\mbox{tr}\left(\F _2^{n+1}\right)-\mbox{tr}\left(\F _1^{n+1}\right) =
(d+\delta )\omega _{2n+1}(\alpha _2 ,\alpha _1)
\end{equation}
for
\begin{equation}
\label{eq:OM}
\omega _{2n+1}(\alpha _2 ,\alpha _1)=(n+1)\int _0^1dt\mbox{tr}\left( 
(\alpha _2 -\alpha _1)\F _t^n\right)
\end{equation}
and $\F _t$ the curvature of $(1-t)\alpha _1 +t\alpha _2$, holds for 
any connections $\alpha _1,\alpha _2$ with curvatures $\F _1, \F _2$. 
Using this in eq. \Ref{eq:CU} gives the following expression for the 
connection of the determinant line bundle:
\begin{equation}
\label{eq:CO}
-2\pi i \frac{1}{(n+1)!}\left(\frac{i}{2\pi }\right) ^{n+1}\int _M 
\omega _{2n+1}(A+(d_A^\ast d_A)^{-1}d_A^\ast ,0).
\end{equation}

The (infinitesimal) consistent anomaly is the variation of the 
effective action under gauge transformations. Thus, it is the negative 
of the restriction of \Ref{eq:CO} to gauge directions, i.e. the fibre 
directions of $\A\rightarrow \A /\G $. Along such directions, 
$\delta $ becomes the BRST operator and $(d_A^\ast d_A)^{-1}d_A^\ast $ 
becomes the ghost $v$. Thus, the consistent anomaly is 
$
c_n\int _M \omega _{2n+1}(A+v,0) 
$
, with $c_n=-\frac{1}{(n+1)!}\left(\frac{i}{2\pi }\right) ^n$. Since $M$
is $2n$-dimensional, it is only the term with one ghost in the expansion
of $\omega _{2n+1}(A+v,0)$ that will give a contribution to the
anomaly. We let $\omega _{2n+1-k}^k(A+v,0)$ denote the part of 
$\omega _{2n+1}(A+v,0)$ that contains $k$ number of ghosts. Then 
$c_n\omega _{2n}^1(A+v,0)$ is the non-integrated anomaly. It is 
well-known, and explicitly proven in \cite{CMM}, that $c_n\int _M
\omega _{2n-1}^2(A+v,0)$ is the Schwinger term. In this case $M$ is to 
be interpreted as the odd-dimensional physical space at a fixed time. 
The forms $\omega _{2n+1-k}^k(A+v,0)$ can be computed by use of eq. 
\Ref{eq:OM}. In 1+1 and 3+1 dimensions it gives the following result 
for the consistent anomaly and Schwinger term:
\begin{eqnarray}
\label{eq:CAS}
c_1\omega _{1}^2(A+v,0)  & = & c_1\mbox{tr}(vdA)  \nonumber \\ 
c_1\omega _{2}^1(A+v,0)  & = & -c_1\mbox{tr}\left( v^2A\right)  \nonumber \\
c_2\omega _{1}^4(A+v,0)  & = & c_2\mbox{tr}\left( vd\left(
AdA+A^3/2\right) \right)  \nonumber \\
c_2\omega _{2}^3(A+v,0)  & = & -c_2\mbox{tr}\left( \left( 
v^2A +vAv+Av^2\right) dA+v^2A^3\right) /2 .
\end{eqnarray} 
If the freedom is used to change the forms $\omega _{2n+1-k}^k(A+v,0)$
by cohomologically trivial terms (i.e. by coboundaries), then these
forms can be given by the following compact expressions that were
first derived by Zumino in \cite{ZU}: 
\begin{eqnarray}
\label{eq:ZUA}
&&\omega _{2n+1-k}^k(A+v,0)  \nonumber \\
&\sim & 
(n+1) \left( \begin{array}{c} n\\ k\end{array}\right) 
\int _0^1dt\,\,(1-t)^k\mbox{str}\left( (dv)^k ,A, \left( 
tdA+t^2A^2\right) ^{n-k}\right)
\end{eqnarray}  
when $0\leq k\leq n$ and
\begin{eqnarray}
\label{eq:ZUB}
\omega _{2n+1-k}^k(A+v,0)  & \sim & 
(-1)^{k-n-1} \left( \begin{array}{c} n\\ k-n-1 \end{array}\right) 
\left(\left( \begin{array}{c} k\nonumber \\ k-n-1\end{array}\right)
\right) ^{-1} \nonumber \\ 
&&\times \mbox{str}\left(v, (v^2)^{k-n-1} ,\left( dv\right) ^{2n-k+1}\right)  
\end{eqnarray}
when $n+1\leq k\leq 2n+1$. Here str means the symmetrized trace and 
$\sim$ means equality up to a coboundary.

Above, we used eq. \Ref{eq:FO} for $\alpha _2=A+(d_A^\ast d_A)^{-1}
d_A^\ast $ and $\alpha _1=0$. When $P$ is a non-trivial bundle it is 
no longer possible to let $\alpha _1$ be zero. Instead, we let it be 
some fixed connection $A_0$ on $P$ (which can be identified with a 
connection on $P\times\A$). By dimensional reasons, this does not 
change eq. \Ref{eq:CU}. The consistent anomaly and Schwinger term with 
such a background connection can be computed in a similar way as above, 
one just uses $\omega _{2n+1-k}^k(A+v,A_0)$ instead. Since the expressions
corresponding to eq. \Ref{eq:ZUA} and eq. \Ref{eq:ZUB} are long and 
not particularly illuminating we shall not present them here (parts of it 
can be found in \cite{Ek1}). The ideas behind the background connection 
are completely analogous with the case without a background. For 
example, they are consistent, but not gauge covariant. To obtain 
covariance, we choose as a background the field itself. We are 
then interested in the (non-consistent) terms coming from the 
expansion of $\omega _{2n+1-k}^k(A+v,A)$ in various ghost degrees. 
With use of eq. \Ref{eq:OM}, the following expression was obtained 
in \cite{Ek1}:
\begin{eqnarray}
\label{eq:IGD}
\omega _{2n+1-k}^k(A+v,A) & = & \sum _{j=0}^{[(k-1)/2]} \frac{n+1}{k-j}
\left( \begin{array}{c} n-j\\ k-2j-1 \end{array}\right) \left( 
\begin{array}{c} n\\ j \end{array}\right)\left(\left( 
\begin{array}{c} k\\ j \end{array}\right)\right) ^{-1}\nonumber \\
&&\times \mbox{str}\left( v, (\delta v )^j,(\delta A)^{k-2j-1}, 
F^{n-k+j+1}\right),
\end{eqnarray}
where a negative power on a factor means that the corresponding term 
is absent in the sum. Recall that $[(k-1)/2]$ is $(k-1)/2$ if $k$ is 
odd and $(k-2)/2$ if $k$ is even. The terms  
\begin{eqnarray}
c_n\omega _{2n}^1(A+v,A) & = & c_n(n+1)\mbox{tr}(vF^n)\nonumber \\
c_n\omega _{2n-1}^2(A+v,A) & = & c_n\frac{n(n+1)}{2}\mbox{str}(v,
\delta A,F^{n-1})
\end{eqnarray}
are the non-integrated covariant anomaly and Schwinger term. 

Let us summarize the results so far in the case of 1+1 and 3+1 
dimensions in Tables 1 and 2, respectively
(we use eq. \Ref{eq:ZUA} and \Ref{eq:ZUB} for the consistent 
formalism) 

\begin{center}
\hspace{-4.2mm}\begin{tabular}{||l|c|c||} \hline
$n=1$ & Anomaly & Schwinger term \\ \hline
\hspace{-2mm}\begin{tabular}{l}\\[-3mm] \\ Consistent \\  \\[1mm]
\end{tabular}
& $c_1\int _M\mbox{tr}((dv)A)$ & $c_1\int _M\mbox{tr}(vdv) $\\\hline
\hspace{-2mm}\begin{tabular}{l}\\[-3mm] \\ Covariant \\ \\[1mm]
\end{tabular} & $\begin{array}{c} c_1\cdot 2\int _M\mbox{tr}\left( 
v\left( dA+A^2\right)\right) \end{array}$ & $-c_1\int _M\mbox{str}( 
v( dv+2vA)) \hspace{1.4mm}$ \\ \hline
\end{tabular}
\\
\vspace{3mm}
Table 1
\end{center}

\begin{center}
\hspace{-4.2mm}\begin{tabular}{||l|c|c||} \hline
$n=2$ & Anomaly & Schwinger term \\ \hline
\hspace{-2mm}\begin{tabular}{l}\\[-3mm] \\ Consistent  \\ \\[1mm]
\end{tabular}
& $\begin{array}{c} c_2\int _M\mbox{tr}\big( (dv)AdA\\  +\frac{1}{2}
(dv)A^3\big)\end{array}\hspace{4mm} $ & $c_2\int _M\mbox{tr}\left( 
(dv)^2A\right)$\\ \hline
\hspace{-2mm}\begin{tabular}{l}\\[-3mm] \\ Covariant \\ \\[1mm]
\end{tabular} & $\begin{array}{c} c_2\cdot 3\int _M\mbox{tr}\\ \left( 
v\left( dA+A^2\right) ^2\right) \end{array}\hspace{0mm}$ & 
$\begin{array}{c} c_2\cdot (-3)\int _M\mbox{str}\big( v\big( dv+vA\\ 
+Av \big)\big( dA+A^2\big)\big)\end{array} \hspace{0mm}$ \\ \hline
\end{tabular}
\\
\vspace{3mm}
Table 2
\end{center}

We shall now evaluate these forms on (anti-hermitean) infinitesimal 
gauge transformations $X,Y\in \mbox{Lie}\G$. Let us do this explicitely 
for the consistent Schwinger term when $n=2$:
\begin{eqnarray}
&&c_2\int _M\mbox{tr}\left( (dv)^2A\right) (X,Y)  =  -c_2
\int _M\mbox{tr}( \partial _iv\wedge\partial _jvA_k)\epsilon^{ijk}
d^3x(X,Y)\nonumber \\
& = & -c_2\int _M\mbox{tr}( (\partial _iX\partial _jY-\partial _iY
\partial _jX)A_k)\epsilon ^{ijk}d^3x.
\end{eqnarray}
The corresponding evaluation of the other forms for $n=1$ and $n=2$
is listed in Tables 3 and 4, respectively.

\begin{center}
\hspace{-4.2mm}\begin{tabular}{||l|c|c||} \hline
$n=1$ & Anomaly & Schwinger term \\ \hline
\hspace{-2mm}\begin{tabular}{l}\\[-3mm] \\ Consistent \\  \\[1mm]
\end{tabular} & $ -c_1\mbox{tr} ((\partial _\mu X) A_\nu) 
\epsilon ^{\mu\nu } \hspace{6mm}$ & $-2c_1\mbox{tr}\left( X\partial _x 
Y\right) $\\ \hline
\hspace{-2mm}\begin{tabular}{l}\\[-3mm] \\ Covariant \\  \\[1mm]
\end{tabular} & $\begin{array}{c} 2c_1\mbox{tr}( X( \partial _\mu A_\nu
\\+A_\mu A_\nu )) \epsilon ^{\mu\nu } \end{array}$ & $\begin{array}{c} 
2c_1\mbox{tr}( X\partial _xY  -[X,Y]A_x)\end{array} \hspace{2mm}$ \\ \hline
\end{tabular}
\\
\vspace{3mm}
Table 3
\end{center}

\begin{center}
\hspace{-4.2mm}\begin{tabular}{||l|c|c||} \hline
$n=2$ & Anomaly & Schwinger term \\ \hline
\hspace{-2mm}\begin{tabular}{l}\\[-3mm] \\ Consistent \\  \\[1mm]
\end{tabular}
& $\begin{array}{c} -c_2\mbox{tr}((\partial _\mu X)(A_\nu
\partial_\rho A_\lambda  \\+\frac{1}{2}A_\nu A_\rho A_\lambda )) 
\epsilon ^{\mu \nu\rho\lambda} \end{array}\hspace{0mm}$ & 
$\begin{array}{c} -c_2\mbox{tr}((\partial _iX\partial _jY 
\\-\partial _iY\partial _jX)  A_k ) \epsilon ^{ijk} \end{array}
\hspace{0mm}$\\ \hline
\hspace{-2mm} Covariant  & $\begin{array}{c}3c_2\mbox{tr}( X(
\partial _\mu A_\nu \\+A_\mu A_\nu ))(\partial _\rho A_\lambda \\ 
+A_\rho A_\lambda )\epsilon ^{\mu \nu\rho\lambda}\end{array}$ &  
$\begin{array}{c}3c_2\mbox{tr}(( X\partial _iY-Y\partial _iX - \\ 
-\left[ X,Y\right] A_i+XA_iY\\ -YA_iX) (\partial _jA_k +A_jA_k)
) \epsilon ^{ijk} \end{array} $ \\ \hline
\end{tabular}
\\
\vspace{3mm}
Table 4
\end{center}


That the covariant anomaly and Schwinger term can be computed by 
expansion of $\omega _{2n+1}(A+v,A)$ was discovered by Kelnhofer \cite{Keln1}. 
An alternative computational scheme leading to covariant cochains differing
from the ones of Kelnhofer was 
given by Tsutsui \cite{Tsu1}. 
To review his approach we reconsider eq. \Ref{eq:OM} 
for $\alpha _2=A+v$ and $\alpha _1=0$. We can then view $\omega _{2n+1}$
as a function $\omega _{2n+1}(A+v|\F )$ of $A+v$ and $\F =(d+\delta )
(A+v)+(A+v)^2$:
\begin{equation}
\omega _{2n+1}(A+v|\F )=(n+1)\int _0^1dt\mbox{tr}\left( (A+v)
\F _t^n\right) , \quad \F _t=t\F +(t^2-t)(A+v).
\end{equation}
 The covariance is broken by the operator $\delta $ in the expression 
for $\F$. Thus, $\omega _{2n+1}(A+v|\F ^\prime )$, with $\F ^\prime 
=d(A+v)+(A+v)^2$, produces covariant terms. This is exactly the same 
terms as the ones appearing in Tsutsui's anti-BRST approach 
\cite{Ek1,Adam2}. In \cite{Ek1} (see \cite{Ek2} for $k=2$) the following 
formula was given for the terms with a given ghost degree:
\begin{equation}
\omega _{2n+1-k}^k(A+v|\F ^\prime )=\frac{n+1}{k}\mbox{tr}\left( 
v(F-\delta (A+v))^n\right) _k,
\end{equation}
where the index $k$ on the right hand side means the part of the 
expression that has $k$ number of ghosts. Comparison with eq. 
\Ref{eq:IGD} reveals that this formula gives the same covariant 
anomaly but the covariant Schwinger term differs by a sign. The 
higher terms seem to be unrelated. This brings us to the question 
of who is right: Kelnhofer or Tsutsui? The formula of Tsutsui seems 
to be motivated by nothing else than covariance. Kelnhofer's formula, 
on the other hand, seems to appear in a natural way: it is obtained 
by putting the background field equal to the field under consideration. 
In the computation of the Schwinger term from determinant line bundles 
for manifolds with boundary, one extends space to a cylindrical 
space-time, \cite{CMM}. On one side of the cylinder one computes the 
Schwinger term by comparison of a fixed vacuum bundle (with respect to 
a background connection) on the other side of the cylinder. In this 
approach it is certainly possible to put the background field equal to 
the field itself, see \cite{Ek3} for details. This clearly defines a 
covariant Schwinger term in a natural way, suggesting that Kelnhofer's 
approach is the correct one. This geometrical approach would not have 
been possible with Tsutsui's result. This explains the importance of 
the sign of the covariant Schwinger term. In
the forthcoming sections we shall demonstrate that indeed Kelnhofer's
result for the covariant Schwinger term is reproduced by quantum
field theoretic computations.

\section{Calculations in 1+1 dimensions}

\subsection{Calculation of Adam}

In this section we want to briefly review the calculation of the
consistent and covariant Schwinger term that was performed in 
\cite{Adam1} for the Abelian case (the chiral Schwinger model). The
generalization to the non-Abelian case is straight-forward and shall be
displayed below, as well. In \cite{Adam1} the Hamiltonian formulation was
used (therefore space-time is 1+1 dimensional Minkowski space),
and the computation started from the second-quantized chiral
fermion field operator in the interaction picture. For fermionic field
operators the Dirac vacuum has to be introduced and operator products
have to be normal-ordered w.r.t. the Dirac vacuum. For the introduction
of the Dirac vacuum the Hilbert space of fermionic states is split into
a positive and negative momentum sub-space (for chiral fermions in two
dimensions energy equals momentum). For the negative momentum sub-space
the role of creation and annihilation operators is then exchanged. At
this point there are two possibilities to split. Either one may split
w.r.t. eigenvalues of the free momentum operator $-i\partial_{x^1}$
and perform normal-ordering (denoted by $N$) for this Dirac vacuum.
A well-known consequence of this normal-ordering is the fact that the
current commutators acquire a central extension (Schwinger term). For
a fermion of positive chirality (where the current 
obeys $J^0 =J^1 =:J$), the
Schwinger term is
\beq
[NJ(x^0 ,x^1),NJ(x^0 ,y^1)] = -\frac{i}{2\pi}\delta '(x^1 - y^1)
\eeq
(here the prime denotes derivative w.r.t. the argument).
The second possibility is to split w.r.t. eigenvalues of the kinetic
momentum operator $-i\partial_{x^1} + eA_1$. Again, a corresponding
Dirac vacuum and normal ordering (denoted by $\wt N$) may be introduced.
It turns out that the kinetically normal-ordered current is related to
the conventionally normal-ordered current in a simple fashion
\cite{IST1,Adam1} 
\beq
\wt NJ(x) = NJ(x) +\frac{e}{2\pi}A_1 (x)
\eeq
therefore $\wt NJ$ has the same commutator (14) as $NJ$. It was
proven in \cite{Adam1} that $NJ$ is the consistent current operator and 
$\wt NJ$ is the covariant current operator.
Now it is very easy to compute the consistent and covariant Gauss law 
commutators. The Gauss law operators are defined as 
($\partial_{x^1}\equiv \partial_1$)
\beq
G(x) =\partial_1 \frac{\delta}{e\delta A_1 (x)} -i NJ(x)
\eeq
\beq
\wt G(x) =\partial_1 \frac{\delta}{e\delta A_1 (x)} -i \wt NJ(x)
\eeq
Here $A_1 (x)$ is treated as a function of space only and the time
variable $x^0$ as a parameter, i.e. $(\delta / \delta A_1 (x^0 ,x^1))
A_1 (x^0 ,y^1)=\delta (x^1 -y^1)$.
The consistent Gauss law commutator is determined by the current
commutator (14),
\beq
[G(x^0 ,x^1),G(x^0 ,y^1)]=\frac{i}{2\pi}\delta '(x^1 -y^1)
\eeq
whereas for the covariant case the functional derivatives contribute, as
well,
\beq
[\wt G(x^0 ,x^1),\wt G(x^0 ,y^1)]=-\frac{i}{2\pi}\delta '(x^1 -y^1).
\eeq
Therefore, the covariant Schwinger term is minus the consistent one,
(18). This relative minus sign is precisely as in Table 1. Observe that
the covariant current is indeed gauge invariant, $[\wt G(x^0 ,x^1),\wt N
J(x^0 ,y^1)] =0$, as it must be. In fact, the relative minus sign
between the consistent and covariant Schwinger terms is a consequence of
this gauge invariance of $\wt NJ$, and therefore independent of all
possible conventions. 

A generalization of the above results to the nonabelian case is straight
forward. The two versions of normal-ordering are defined as in the
abelian case, and they lead to the same relation as in (15), up to an
additional colour index
\beq
\wt NJ^a(x) = NJ^a(x) +\frac{e}{2\pi}A^a_1 (x).
\eeq
Further, the current commutator acquires a canonical piece as well,
\beq
[NJ^a (x^0 ,x^1),NJ^b (x^0 ,y^1)] = -if^{abc} NJ^c (x^0 ,x^1) \delta (x^1
- y^1)-\frac{i}{2\pi}\delta^{ab} \delta '(x^1 - y^1)
\eeq
(for the commutator $[\wt NJ^a (x),\wt NJ^b (y)]$, the same expression
is obtained, again with $NJ^c$ on the r.h.s., {\em not}
$\wt NJ^c$, as is obvious from (20)). 
The generator of
time-independent gauge transformations on gauge fields,
\beq
\delta^a (x):=(\delta^{ab}\partial_1 +ef^{acb}A^c_1
(x))\frac{\delta}{e\delta A^b_1 (x)} 
\eeq
obeys the commutation relation
\beq
[\delta^a (x^0,x^1),\delta^b (x^0,y^1)]=-f^{abc}\delta^c (x^0,x^1)\delta
(x^1 - y^1).
\eeq
The consistent and covariant Gauss law operators are defined as
\beq
G^a (x)=\delta^a (x) -iNJ^a (x)
\eeq
and
\beq
\wt G^a (x) = \delta^a (x) -i\wt NJ^a (x)
\eeq
respectively.
Their anomalous commutators may be easily computed,
\beq
[G^a (x^0 ,x^1),G^b (x^0 ,y^1)] + f^{abc}G^c (x^0 ,x^1)\delta(x^1 -y^1)=
\frac{i}{2\pi}\delta^{ab}\delta' (x^1 -y^1)
\eeq
\bdi
[\wt G^a (x^0 ,x^1),\wt G^b (x^0 ,y^1)] + f^{abc}\wt G^c (x^0 ,x^1)
\delta(x^1 -y^1)= 
\edi
\beq
-\frac{i}{2\pi}\delta^{ab}\delta' (x^1 -y^1)
+\frac{i}{2\pi} f^{abc}A^c (x^0 ,x^1)\delta (x^1 -y^1) .
\eeq
As in the abelian case, 
the anomalous commutators agree with the ones in Table 1, and
again this is most easily seen for the relative minus sign of the $\delta'
(x^1 -y^1)$ term. This relative sign may be related to the fact
that the covariant current has to transform covariantly under a gauge
transformation (i.e., the $\delta' $ terms must cancel)
\beq
[\wt G^a (x^0 ,x^1),\wt NJ^b (x^0 ,y^1)]=-f^{abc}\wt NJ^c (x^0
,x^1)\delta (x^1 -y^1)
\eeq
as may be checked easily.

\subsection{Calculation of Hosono and Seo}

In this section we shall use the Hosono and Seo approach \cite{HS}
for the calculation of the equal-time commutators of the covariant and
consistent Gauss law operator. The calculation is performed in Minkowski
space, $g_{\mu\nu}={\rm diag}\, (1,-1)$, $\varepsilon_{01}=1$,
with the gamma matrices obeying
the usual Clifford algebra relation
$\gamma ^{\mu }\gamma ^{\nu }+\gamma ^{\nu }\gamma ^{\mu }=2g^{\mu \nu
  }$, and $\gamma _{5}=\gamma ^{0}\gamma ^{1}$.
The anti-Hermitian matrices $t^{i}$ are the generators of a non-abelian 
algebra $\left[ t^{a},t^{b}\right] =f^{abc}t^{c}$,
and we denote $A_{k}=A_{k}^{a}t^{a}$.

The Hamiltonian of the chiral fermion interacting with an external gauge
potential is
\begin{equation}
{\mathcal H}\left( A\right) =-i\int dx\,\left[ \,\bar{\psi}\left(t,
x\right)
\gamma ^{1}\frac{1+\gamma _{5}}{2}\left( \partial _{1}+A_{1}^{a}\left(
t,x\right) t^{a}\right) \psi \left(t, x\right) \right] 
\label{Hamiltonian}
\end{equation}
where we chose the Weyl gauge $\left( A^{0}\left( t,x\right) =0\right)$. 

We expand the Fermion field as 
\beq
\psi(t,x) =\sum_n \alpha_n (t)  \zeta_n (t,x), 
\eeq
where
$\zeta_n (t,x)$ are  eigenfunctions of the full Hamiltonian (29) 
with eigenvalues
$E_n (t)$. In the quantized theory the $\alpha_n$ are treated as operators
satisfying the canonical anticommutation relation
\begin{equation}
\left\{ \alpha _{n,}\alpha _{m}^{+}\right\}  =\delta_{nm}
\end{equation}
and the Dirac vacuum is defined as
\begin{eqnarray}
\alpha _{n}\left( t\right) \left| 0,A\left( t\right) \right\rangle _{S}
&=&0,\quad E_{n}\left( t\right) >0,  \nonumber \\
\alpha _{n}^{+}\left( t\right) \left| 0,A\left( t\right) \right\rangle_{S}
&=&0,\quad E_{n}\left( t\right) <0.
\end{eqnarray}
Observe that the expansion of the Fermion field operators w.r.t. the
eigenfunctions of the full Hamiltonian (29) automatically implies that we
use the Schroedinger picture.

Singular operator products are regularized in \cite{HS} by an
exponential damping of high frequencies. The regularized current reads
\bdi
\left( j^{\mu a}\left( x\right) \right) _{reg}=\sum_{n,m}\alpha
_{n}^{+}\left( t\right) \zeta _{n}^{+}\left( t,x\right) e^{-\left(
\varepsilon /2\right) E_{n}^{2}\left( t\right) }\gamma ^{0}\gamma ^{\mu
}%
\frac{1+\gamma _{5}}{2}e^{-\left( \varepsilon /2\right) E_{m}^{2}\left(
t\right) }\zeta _{m}\left( t,x\right) \alpha _{m}\left( t\right) 
\edi
\beq
=\sum_{n,m}\alpha _{n}^{+}\left( t\right) \zeta _{n}^{+}\left(
t,x\right)
e^{-\left( \varepsilon /2\right) E_{n}^{2}\left( t\right)}
t^{a}e_{m}^{-\left( \varepsilon /2\right) E_{m}^{2}\left( t\right) }
\zeta_{m}\left( t,x\right) \alpha _{m}\left( t\right) 
\eeq
(where $j^0 = j^1$ was used in the second line, which holds for the
chiral current (33)). 
The current in (33) is regularized covariantly, therefore it will lead to
the covariant anomaly and Schwinger term. The consistent current $J^\mu$
is obtained by adding the Bardeen--Zumino polynomial $\Delta j^\mu$,
 \begin{equation}
J^{\mu }\left( x\right) =j^{\mu }\left( x\right) +\Delta j^{\mu }
\left(x\right) ,
\end{equation}
\begin{equation}
\Delta j^{\mu }\left( x\right) =-\frac{i}{4\pi }t^{a}\varepsilon ^{\mu
  \nu }
{\mathrm tr}\left( t^{a}A_{\nu }\right) .
\end{equation}

These currents lead to the covariant and consistent anomalies
\begin{equation}
{\mathcal A}_{cov}^a\left( x\right) =-\left( D^\mu \langle j_{\mu }
\rangle \right)^a\left(
x\right) =\frac i{2\pi }\varepsilon _{\mu \nu }{\mathrm tr}\left(
t^a\left(
\partial ^\mu A^\nu +A^\mu A^\nu \right) \right) \left( x\right)
\label{covariant anomaly}
\end{equation}
and
\begin{equation}
{\mathcal A}_{con}^a\left( x\right) =-\left( D^\mu \langle J_{\mu }
\rangle \right)^a\left( x\right) =
\frac i{4\pi }\varepsilon _{\mu \nu }
{\mathrm tr}\, t^a\partial ^\mu A^\nu \left( x\right) .
\label{consistent anomaly}
\end{equation}

The covariant ($\tilde G^a$) and consistent ($G^a$) Gauss law operators
read
\begin{equation}
\tilde{G}^{a}\left( x\right) =X^{a}\left( x\right) +j^{0a}\left(
x\right) 
\label{covariant Gauss Law}
\end{equation}
\begin{equation}
G^{a}\left( x\right) =X^{a}\left( x\right) +J^{0a}\left( x\right) 
\label{consistent Gauss law}
\end{equation}
where
\begin{equation}
X^{a}\left( x\right) =-\left( \partial _{1 }\frac{\delta }{\delta
  A_1^{
a}\left( x\right) }+f^{abc}A_1^{ b}\left( x\right) \frac{\delta
}{\delta
A_1^{c}\left( x\right) }\right)
\end{equation}
generates time-independent gauge transformations of the external gauge
field.

Assuming that the non-canonical parts ($n.c.$) of the commutator of the
covariant and consistent Gauss laws 
are $c$-numbers it is sufficient to consider their vacuum expectation
values (VEVs) only. The calculation in the Hosono and Seo approach is
rather lengthy, therefore it is performed in the Appendices A -- C. Here we
just present the final form of the covariant Schwinger term
\begin{eqnarray}
\widetilde{ST}^{ab} &=&\left\langle \left[ \tilde{G}^{a}\left( x\right)
,
\tilde{G}^{b}\left( y\right) \right] _{n.c.}\right\rangle =-\left\langle 
\left[ j^{0a}\left( x\right) ,j^{0b}\left( y\right) \right]_{n.c.}
\right\rangle =  \nonumber \\
&=&\frac{i}{2\pi }\partial _{x}\delta (x-y)\cdot {\mathrm tr}t^{a}t^{b}+\frac{
i}{2\pi }\delta (x-y)\cdot {\mathrm tr}t^{a}\left[ A_{1}\left( y\right) ,t^{b}
\right] ,  \label{covariant ST}
\end{eqnarray}
and the consistent one
\begin{equation}
ST^{ab}=\left\langle \left[ G^{a}\left( x\right) ,G^{b}\left( y\right) 
\right] _{n.c.}\right\rangle =\frac{i}{4\pi }\delta \left(
x-y\right)
\cdot {\mathrm tr}\left( \left[ t^{a},A_{1}\right] t^{b}\right) .
\label{consistent ST}
\end{equation}
Comparing the results (41) and (42) with the expressions for the 1+1 dim
Schwinger terms in (6) (for the consistent case) and Table 1 (for the
covariant case), we find that these terms agree.
Therefore, the method of Hosono and Seo reproduces the result of
Kelnhofer \cite{Keln1}.

\section{Method of Wess}

In this section we want to review the papers of  Schwiebert
\cite{Schw1} and  Kelnhofer \cite{Keln2} who used the method of
Wess \cite{Wess1} for the calculation of the consistent \cite{Schw1}
and covariant \cite{Keln2} Schwinger terms (ST), respectively. The
central idea of this method is to infer the current commutators from the
time derivatives of a (time-ordered) current two-point function, by 
using the general relation $\partial^0_{x} TA(x)B(y) = \delta(x^0 -y^0)
[A(x), B(y)]$. As the anomaly is a (covariant) derivative of the current
VEV (one-point function), and further current insertions are obtained by
functional derivatives w.r.t. the external gauge potential $A^\mu_a$, 
the current commutator may be related to a functional
derivative of the anomaly.

The authors of \cite{Schw1} and \cite{Keln2} used slightly different
conventions. For our purposes it is important to have the same
conventions for both the consistent and covariant cases, because we
want to determine one relative sign. Therefore we shall repeat the major
steps in the calculations of \cite{Schw1} and \cite{Keln2} within
our specific set of conventions. We choose anti-Hermitean Lie algebra
generators $\lambda_a$,
\beq
[\lambda_a ,\lambda_b ]=f_{abc}\lambda_c
\eeq
where $f_{abc}$ are the structure constants. Further we choose Euclidean
conventions in this section ($g^{\mu\nu}=\delta^{\mu\nu}$), mainly
because the path integral computation of both the consistent \cite{AGG1}
and covariant \cite{Fuji1} anomaly was done in Euclidean space as well
(for our conventions see e.g. \cite{Bertl1}). ``Space-time'' indices
(running from 0 to 1 in $d=2$ and from 0 to 3 in $d=4$) are denoted by
Greek letters $\mu ,\nu , \ldots$ and pure space indices are denoted by
latin letters $ k,l,m$. For the Ward operator we choose
\beq
X_a (x) =-(D^\mu_x)_{ab}\frac{\delta}{\delta A^\mu_b (x)}\equiv
-(\delta_{ab}\partial^\mu_x +f_{acb}A^\mu_c (x))
\frac{\delta}{\delta A^\mu_b (x)}
\eeq
\beq
[X_a (x),X_b (y)]=f_{abc}X_c (x)\delta(x-y) .
\eeq
The Euclidean vacuum functional is
\beq
Z[A] =e^{-W[A]}=\langle 0|T^* e^{-\int dx \wh J^\mu_a (x) 
A^\mu_a (x) }|0\rangle 
\eeq
where $A^\mu_a$ is the external gauge potential and $\wh J^\mu_a$ is
a covariantly regularized current operator, which necessarily depends on
$A^\mu_a$ for an anomalous gauge theory. Further $T^*$ is the
Lorentz covariantized time-ordered product that results from covariant
perturbation theory. 

\subsection{Consistent case}

For the VEV of the consistent current $J^\mu_a$ (one-point function)
we have ($\int \wh J A
\equiv \int dx \wh J^\mu_a (x)A^\mu_a (x)$)
\bdi
\langle 0|T^* J^\mu_a (x)e^{-\int \wh JA}|0\rangle e^W :=\frac{\delta W}{\delta
A^\mu_a (x)} 
\edi
\beq
= \langle 0|T^* (\wh J^\mu_a (x) +\int dy \frac{\delta \wh J^\lambda_b (y)}{
\delta A^\mu_a (x)}A^\lambda_b (y))e^{-\int \wh JA}|0\rangle e^W
\eeq
and for the two-point function we get
\bdi
\frac{\delta^2 W}{\delta A^\mu_a (x) \delta A^\nu_b (y)} = - \langle 0|T^*
J^\mu_a (x) J^\nu_b (y) e^{-\int \wh JA}|0\rangle  e^W
\edi
\beq
+\langle 0|T^* \frac{\delta J^\mu_a (x)}{\delta A^\nu_b (y)}e^{-\int \wh JA}|0\rangle 
e^W + \frac{\delta W}{\delta A^\mu_a (x)}\frac{\delta W}{\delta A^\nu_b (y)}
\eeq
\beq
=: -T^{*\mu\nu}_{ab}(x,y) + \Theta^{\mu\nu}_{ab}(y)\delta (x-y) +\cdots
\eeq
where in (49) we have defined abbreviations for 
the first and second term of (48) and indicated the
third (disconnected) term by ellipses. Here it is assumed that $J^\mu_a$
depends on $A^\mu_a$ only in a local fashion \cite{Schw1}.

Now we should re-express the $T^*$ product by the ordinary $T$ product
that is defined via $\theta$ functions. For the zero- and one-point
functions we may simply define the $T^*$ product by the $T$ product, 
because the latter leads to Lorentz-covariant expressions. On the other
hand, for the two-point function $\langle T^* J(x)J(y)\rangle $ there occurs a
difference (seagull term $\tau^{\mu\nu}_{ab}$) 
at coinciding space-time points, and this
seagull term is proportional to $\delta (x-y)$ \cite{GrJa1,Jack1}. 
Denoting the ordinary $T$ product by $T^{\mu\nu}_{ab}(x,y)$, we have
\beq
T^{*\mu\nu}_{ab}(x,y)=T^{\mu\nu}_{ab}(x,y) +
\tau^{\mu\nu}_{ab}(y)\delta(x-y).
\eeq
For the divergence of $T^{\mu\nu}_{ab}$ we get, using the definition of
the $T$ product,
\bdi
\partial^\mu_x T^{\mu\nu}_{ab}(x,y)= \partial^\mu_x \Bigl( \theta (x^0
-y^0) \langle 0|(Te^{-\int_{x^0}^\infty \wh JA})J^\mu_a
(x)(Te^{-\int_{y^0}^{x^0} \wh JA})J^\nu_b (y)(Te^{-\int_{-\infty}^{y^0}
  \wh JA})|0\rangle 
\edi
\bdi
+((\mu ,a,x) \leftrightarrow (\nu ,b,y)) \Bigr) e^W
\edi
\bdi
= \delta (x^0 -y^0)\langle 0|T[J^0_a (x),J^\nu_b (y)]e^{-\int \wh JA}|0\rangle e^W
+ \langle 0|T\partial^\mu_x J^\mu_a (x)J^\nu_b (y) e^{-\int \wh JA}|0\rangle e^W
\edi
\beq
-\langle 0|T[J^0_a (x),\int_{z^0 = x^0} {\bf d}z \wh J^\lambda_c (z)A^\lambda_c
(z)]J^\nu_b (y) e^{-\int \wh JA}|0\rangle e^W
\eeq
where ${\bf d}z$ is w.r.t. the spacial coordinates only. The term
containing $\partial^\mu_x J^\mu_a (x)$ does not produce $\delta$
functions and may therefore be neglected. Further, $\wh J$ in the third
term may be replaced by $J$ without introducing $\delta$ function like
contributions. For the commutator we use (in our Euclidean conventions
$J^\nu_b$ is anti-Hermitean)
\beq
\delta(x^0 -y^0)[J^0_a (x),J^\nu_b (y)] = f_{abc}J^\nu_c (y)\delta(x-y)
+C^{0\nu}_{ab}(y)\delta(x-y) + S^{0\nu k}_{ab}(y)\partial^k_x \delta (x-y)
\eeq
Re-inserting this commutator into (51) and omitting disconnected terms we
get
\bdi
\partial^\mu_x T^{\mu\nu}_{ab}(x,y) = C^{0\nu}_{ab}(y)\delta (x-y)
+S^{0\nu k}_{ab}(y)\partial^k_x \delta (x-y)
\edi
\beq
+ f_{abc}\frac{\delta W}{\delta A^\nu_c (y)}\delta(x-y) - f_{adc}
A^\lambda_d (x) T^{\lambda\nu}_{cb}(x,y) .
\eeq
This result has to be related to the functional derivative of the
consistent anomaly, where the consistent anomaly itself is defined as
\beq
{\cal A}_a (x):= X_a (x)W[A].
\eeq
Explicitly, we have in 2 and 4 dimensions ($A_\mu \equiv A^\mu_a
\lambda_a$) 
\beq
d=2\, : \quad {\cal A}_a (x) = c_1 \epsilon^{\mu\nu}\tr \lambda_a
\partial^\mu A^\nu 
\eeq
\beq
d=4\, ,\quad {\cal A}_a (x) = c_2 \epsilon^{\mu\nu\rho\sigma} \tr
\lambda_a \partial^\mu (A^\nu \partial^\rho A^\sigma + \frac{1}{2} A^\nu
A^\rho A^\sigma )
\eeq
where $c_1$ and $c_2$ are some constants. From these explicit
expressions we may express the functional derivatives of the anomalies as
\beq
\frac{\delta {\cal A}_a (x)}{\delta A^\nu_b (y)}= I^{\mu\nu}_{ab}(x)
\partial^\mu_x \delta (x-y) + (\partial^\mu_x I^{\mu\nu}_{ab}(x))
\delta (x-y) = I^{\mu\nu}_{ab}(y)
\partial^\mu_x \delta (x-y)
\eeq
where the last equality follows from properties of the $\delta$
function. Explicitly we have
\beq
d=2\, ,\quad I^{\mu\nu}_{ab}(y) = c_1 \epsilon^{\mu\nu}\tr \lambda_a
\lambda_b
\eeq
\beq
d=4\, ,\quad I^{\mu\nu}_{ab}(y) = \frac{c_2}{2} 
\epsilon^{\mu\nu\rho\sigma} \tr \Bigl( \{\lambda_a , \lambda_b \}
(2\partial^\rho A^\sigma +A^\rho A^\sigma ) - \lambda_a A^\rho
\lambda_b A^\sigma \Bigr) .
\eeq
For later convenience we also note that
\beq
\frac{\delta {\cal A}_b (y)}{\delta A^\mu_a (x)}
=- I^{\nu\mu}_{ba}(y)\partial^\nu_x \delta (x-y) +(\partial^\nu_y 
I^{\nu\mu}_{ba}(y))\delta (x-y) .
\eeq
On the other hand, we may use the definition (54) of the anomaly (and
expression (44) for the Ward operator) to relate the functional
derivative (57) to the two-point function (49). We get
\bdi
\frac{\delta {\cal A}_a (x)}{\delta A^\nu_b (y)}= -f_{abc}\delta (x-y)
\frac{\delta W}{\delta A^\nu_c (x)}
\edi
\bdi
+(\delta_{ac}\partial^\mu_x + f_{adc}A^\mu_d (x))(T^{*\mu\nu}_{cb}(x,y)
-\Theta^{\mu\nu}_{cb}(y)\delta (x-y))
\edi
\bdi
= C^{0\nu}_{ab}(y)\delta (x-y) + S^{0\nu k}_{ab}(y)\partial^k_x \delta
(x-y)
\edi
\beq
+ \sigma^{\mu\nu}_{ab}(y) \partial^\mu_x \delta (x-y) + f_{adc}A^\mu_d
(y) \sigma^{\mu\nu}_{cb}(y)\delta (x-y)
\eeq
where we introduced
\beq
\sigma^{\mu\nu}_{ab}(y):= \tau^{\mu\nu}_{ab}(y) -
\Theta^{\mu\nu}_{ab}(y) .
\eeq
Comparing the coefficients of $\delta (x-y)$, $\partial^k_x \delta
(x-y)$ and $\partial^0_x \delta (x-y)$ in (57) and (61) leads to
\beq
C^{0\nu}_{ab}(y)+ f_{adc}A^\mu_d (y) \sigma^{\mu\nu}_{cb}(y)=0
\eeq
\beq
S^{0\nu k}_{ab}(y) + \sigma^{k\nu}_{ab}(y) = I^{k\nu}_{ab}(y)
\eeq
\beq
\sigma^{0\nu}_{ab}(y)=I^{0\nu}_{ab} (y) .
\eeq
For a determination of $S^{00k}_{ab}$ and $C^{00}_{ab}$ we need
$\sigma^{k0}_{ab}$ about which we have no information yet (here we slightly
deviate from the calculation of \cite{Schw1} and follow the arguments
of \cite{Keln2}, but the final result will agree with the result of
\cite{Schw1} up to the difference in conventions). For this purpose we
compute, analogously to (61),
\bdi
\frac{\delta {\cal A}_b (y)}{\delta A^\mu_a (x)}= -f_{bac}\delta (x-y)
\frac{ \delta W}{\delta A^\mu_a (x)}
\edi
\beq
+(\delta_{bc} \partial^\nu_y + f_{bdc}A^\nu_d (y))(T^{*\mu\nu}_{ac}(x,y)
- \Theta^{\mu\nu}_{ac}(y)\delta(x-y))
\eeq
and use
\bdi
\partial^\nu_y T^{\mu\nu}_{ab}(x,y)= \ldots =-C^{\mu 0}_{ab} (y)\delta
  (x-y) -S^{\mu 0k}_{ab}(y)\partial^k_x \delta (x-y)
\edi
\beq
+ f_{dbc}A^\nu_d (y) T^{\mu\nu}_{ac}(x,y) - f_{abc}\frac{\delta
  W}{\delta A^\mu_c (y)}\delta (x-y)
\eeq
to arrive at
\bdi
\frac{\delta {\cal A}_b (y)}{\delta A^\mu_a (x)}= - C^{\mu 0}_{ab} (y)
\delta (x-y) - S^{\mu 0k}_{ab}(y)\partial^k_x \delta (x-y)
\edi
\beq
+\delta (x-y)(\delta_{bc}\partial^\nu_y +f_{bdc}A^\nu_d (y))\sigma^{\mu
  \nu}_{ ac}(y) - \sigma^{\mu\nu}_{ab} (y)\partial^\nu_x \delta (x-y).
\eeq
Comparison of coefficients of (60) and (68) leads to
\beq
-C^{\mu 0}_{ab}(y) +\partial^\nu_y \sigma^{\mu\nu}_{ab}(y) + f_{bdc}
A^\nu_d (y) \sigma^{\mu\nu}_{ac}(y) = \partial^\nu_y I^{\nu\mu}_{ba}(y)
\eeq
\beq
S^{\mu 0k}_{ab}(y) + \sigma^{\mu k}_{ab}(y) = I^{k\mu}_{ba}(y)
\eeq
\beq
\sigma^{\mu 0}_{ab}(y) = I^{0\mu}_{ba}(y) .
\eeq
Together with (63)--(65) this may be solved for $S^{00k}_{ab}$ and
$C^{00}_{ab}$ 
\beq
S^{00k}_{ab}(y)=I^{k0}_{ab}(y) - I^{0k}_{ba}(y)
\eeq
\beq
C^{00}_{ab}(y)= -f_{adc}A^\mu_d (y)I^{0\mu}_{bc}(y) .
\eeq
In addition we find from (69) and (73) the consistency condition
\beq
\partial^\nu_y (I^{0\nu}_{ab}(y) - I^{\nu 0}_{ba}(y)) + A^\nu_d (y)(
f_{adc} I^{0\nu}_{bc}(y) + f_{bdc}I^{0\nu}_{ac}(y))=0
\eeq
which holds for both 2 and 4 dimensions, as may be checked easily. So
far we have determined the anomalous $[J^0,J^0]$ commutator, see (52),
(72) and (73). We still need the commutator of $J^0_a$ and the Ward operator
$X_b$. As $X_b$ does not contain fermionic degrees, this commutator is
equal to the action of $X_b$ on $J^0_a$,
\beq
X_b (y)J^0_a (x)\equiv \delta (x^0 -y^0) [X_b (y),J^0_a (x)] .
\eeq
This commutator may be inferred from the relation 
\bdi
X_b (y)\frac{\delta W}{\delta A^\mu_a (x)}= -(D^\nu_y)_{bc}
\frac{\delta}{\delta A^\nu_c (y)}\langle 0|TJ^\mu_a (x)e^{-\int \wt JA}|0\rangle e^W
\edi
\beq
=\langle 0|T(X_b (y)J^\mu_a (x))e^{-\int \wt JA}|0\rangle e^W + (D^\nu_y)_{bc} \langle 0|
TJ^\mu_a (x) J^\nu_c (y)e^{-\int \wt JA}|0\rangle e^W .
\eeq
Here we used the fact that in the one-point function (47) the $T^*$
product is equal to the $T$ product. It is important to use the $T$
product here, because we want to extract the (Lorentz non-covariant)
commutator $[J^0_a ,X_b]$ directly, without some covariantizing
seagulls. Now we assume that the commutator (75) contains no fermionic
degrees of freedom, i.e., it may be extracted from the VEV. Using (49)
and (50) for the two-point function we find
\beq
\delta (x^0 -y^0)[X_b (y) ,J^\mu_a (x)] +(D^\nu_y)_{bc}T^{\mu\nu}_{ac}
(x,y) =   (D^\nu_y)_{bc}(T^{\mu\nu}_{ac} + \sigma^{\mu\nu}_{ac}(y)
\delta (x-y))
\eeq
or, for $\mu =0$ and using (65),
\beq
\delta(x^0 -y^0)[J^0_a (x),X_b (y)] = -(D^k_y)_{bc}(I^{0k}_{ac}(y)\delta
(x-y)) .
\eeq
Actually, for the Gauss operator we only need the Ward operator
restricted to purely spacial gauge transformations. In addition it is
preferable to get rid of the time coordinate altogether. Therefore we
define a spacial Ward operator 
\beq
{\bf X}_a (x) := -\int dx^0 (D^k_x)_{ab}\frac{\delta}{\delta A^k_b (x)}
\eeq
\beq
[{\bf X}_a (x),{\bf X}_b (y)] = f_{abc}{\bf X}_c (x){\vec \delta}(x-y)
\eeq
where ${\vec \delta}(x-y)$ is the spacial $\delta$ function. The Gauss
operator is
\beq
G_a (x) = J^0_a (x) + {\bf X}_a (x) .
\eeq
Using (72), (73) and (78) we find for the anomalous part of the commutator
(i.e., the Schwinger term)
\bdi
{\cal G}_{ab}(x,y):= [G_a (x) ,G_b (y)] - f_{abc}G_c (x){\bf
  \delta}(x-y)
\edi
\bdi
= C^{00}_{ab}(y){\vec \delta}(x-y) + S^{00k}_{ab}(y)\partial^k_x {\vec
  \delta} (x-y) +[J^0_a (x),{\bf X}_b (y)] + [{\bf X}_a (x),J^0_b (y)]
\edi
\beq
=-(f_{bdc}A^k_d (y)I^{0k}_{ac}(y) + \partial^k_y I^{0k}_{ab}(y)) {\vec
  \delta} (x-y) .
\eeq
Before evaluating this expression explicitly for $d=2$ and $d=4$, we
want to find the analogous result for the covariant case, following
\cite{Keln2}. 

\subsection{Covariant case}

The VEV of the covariant current $\wt J^\mu_a$ is related to the
consistent one by the Bardeen--Zumino polynomial $\Lambda^\mu_a$,
\beq
\langle 0|T^* \wt J^\mu_a (x)e^{-\int \wh JA}|0\rangle e^W = 
\langle 0|T^*J^\mu_a (x)e^{-\int
  \wh JA}|0\rangle e^W + \Lambda^\mu_a (x) .
\eeq
This leads to the covariant anomaly $\wt {\cal A}_a (x)$,
\beq
\wt {\cal A}_a (x) = -(D^\mu_x)_{ab}
\langle 0|T^* \wt J^\mu_b (x)e^{-\int \wh JA}|0\rangle e^W = {\cal A}_a (x)
-(D^\mu_x)_{ab} \Lambda^\mu_b (x) .
\eeq
Explicitly the covariant anomalies are
\beq
d=2\, ,\quad \wt {\cal A}_a (x) = 2c_1 \epsilon^{\mu\nu}\tr \lambda_a
(\partial^\mu A^\nu + A^\mu A^\nu )
\eeq
\beq
d=4 \, ,\quad \wt {\cal A}_a (x) = 3c_2 \epsilon^{\mu\nu\rho\sigma} \tr
\lambda_a (\partial^\mu A^\nu + A^\mu A^\nu )(\partial^\rho A^\sigma +
A^\rho A^\sigma )
\eeq
where the constants $c_1$, $c_2$ are the {\em same} as in the
consistent case, see (55) and (56). The two-point functions are defined
analogously to (48)--(50) as
\bdi
\frac{\delta}{\delta A^\nu_b (y)}\langle 0|T^* \wt J^\mu_a (x) e^{-\int \wh
  JA}|0\rangle  e^W = - 
\langle 0|T^* \wt J^\mu_a (x) \wt J^\nu_b (y) e^{-\int \wh JA}
|0\rangle  e^W
\edi
\bdi
+\langle 0|T^* \frac{\delta \wt J^\mu_a (x)}{\delta A^\nu_b (y)} e^{-\int \wh
  JA} |0\rangle e^W + \ldots
\edi
\beq
=: - \wt T^{*\mu\nu}_{ab}(x,y) +\wt \Theta^{\mu\nu}_{ab}(y) \delta (x-y)
+ \ldots
\eeq
\beq
=: - \wt T^{\mu\nu}_{ab}(x,y) -\wt\sigma^{\mu\nu}_{ab}(y) \delta (x-y)
+ \ldots
\eeq
where the ellipses denote disconnected terms and all definitions are
analogous to the consistent case. Further, the
computation of $\partial^\mu_x \wt T^{\mu\nu}_{ab}(x,y)$ is completely
analogous to the consistent case, see (51)--(53). Parametrizing the
covariant commutator in an analogous way,
\beq
\delta (x^0 -y^0 )[\wt J^0_a (x),\wt J^\nu_b (y)] = f_{abc}\wt J^\nu_c
(x) \delta (x-y) + \wt C^{0\nu}_{ab} (y) \delta (x-y) + \wt S^{0\nu
    k}_{ab} (y) \partial^k_x \delta (x-y)
\eeq
leads to a result analogous to (53),
\bdi
\partial^\mu_x \wt T^{\mu\nu}_{ab}(x,y)= \wt C^{0\nu}_{ab} (y)\delta
  (x-y)+ \wt S^{0\nu k}_{ab} (y)\partial^k_x \delta (x-y) 
\edi
\beq
+f_{abc} \delta (x-y)\langle 0|T \wt J^\nu_c (y)e^{-\int 
\wh JA}|0\rangle e^W - f_{adc}
  A^\lambda_d (x) \wt T^{\lambda\nu}_{cb} (x,y) .
\eeq
Again, this should be related to the functional derivative of the
(covariant) anomaly. We express this functional derivative as
\beq
\frac{\delta \wt {\cal A}_a (x)}{\delta A^\nu_b (y)}=
\wt I^{\mu\nu}_{ab}(y)\partial^\mu_x \delta (x-y) + \wt B^\nu_{ab} (y)
\delta (x-y)
\eeq
(we do not display the explicit expressions for $\wt I$ and $\wt B$ for
$d=2$ or $d=4$, because we do not need them in the sequel).

On the other hand, using the definition of $\wt {\cal A}_a$, relating
its functional derivative to the two-point function (88) and inserting
(90) for $\partial^\mu \wt T^{\mu\nu}_{ab}$ leads to
\bdi
\frac{\delta \wt {\cal A}_a (x)}{\delta A^\nu_b (y)}= \wt C^{0\nu}_{ab}
  (y)  \delta (x-y) + S^{0\nu k}_{ab}(y) \partial^k_x \delta (x-y)
\edi
\beq
+\wt \sigma^{\mu\nu}_{ab}(y)\partial^\mu_x \delta (x-y) + f_{adc}A^\mu_d
(y) \wt \sigma^{\mu\nu}_{cb}(y)\delta (x-y)
\eeq
and therefore to the equations
\beq
\wt C^{0\nu}_{ab}(y) = \wt B^\nu_{ab} (y) - f_{adc}A^\mu_d (y)\wt
  \sigma^{\mu\nu}_{cb} (y)
\eeq
\beq
\wt S^{0\nu k}_{ab} (y) = \wt I^{k\nu}_{ab} (y) - \wt \sigma^{k\nu}_{ab}
      (y)
\eeq
\beq
\wt \sigma^{0\nu}_{ab} (y) = \wt I^{0\nu}_{ab} (y) .
\eeq
Again we miss information on $\wt \sigma^{k0}_{ab}(y)$, which we may
infer from $(\delta \wt {\cal A}_b (y)/\delta A^\mu_a (x))$. We find
\bdi
\frac{\delta \wt {\cal A}_b (y)}{\delta A^\mu_a (x)}=
-f_{bac}\delta(x-y) \langle 0|
T^* \wt J^\mu_c (y)e^{-\int \wh JA}|0\rangle e^W
\edi
\bdi
-(D^\nu_y)_{bc}\frac{\delta}{\delta A^\mu_a (x)}\langle 0| T^* \wt J^\nu_c (y)
e^{-\int \wh JA}|0\rangle  e^W
\edi
\bdi
=-f_{bac} \delta (x-y) \langle 0|
T^* \wt J^\mu_c (y)e^{-\int \wh JA}|0\rangle e^W
\edi
\beq
-(D^\nu_y)_{bc} \Bigl( \frac{\delta}{\delta A^\nu_c (y)} \langle 0| T^* \wt
J^\mu_a (x) e^{-\int \wh JA}|0\rangle e^W - {\cal F}^{\mu\nu}_{ab}(x,y) \Bigr)
\eeq
\beq
{\cal F}^{\mu\nu}_{ab}(x,y) := \frac{\delta \Lambda^\mu_a(x)}{\delta
A^\nu_b (y)} - \frac{\delta \Lambda^\nu_b (y)}{\delta A^\mu_a (x)}
\eeq
where we used relation (83) between consistent and covariant current VEV
and the commutativity of functional derivatives (see \cite{Keln2}).
Computing $\partial^\nu_y \wt T^{\mu\nu}_{ab}(x,y)$ as in the consistent
case yields
\bdi
\frac{\delta \wt {\cal A}_b (y)}{\delta A^\mu_a (x)} = -\wt C^{\mu
  0}_{ab} (y) \delta (x-y) - \wt S^{\mu 0k}_{ab} (y) \partial^k_x \delta
(x-y) 
\edi
\beq
+ (D^\nu_y)_{bc} {\cal F}^{\mu\nu}_{ac}(x,y) + \delta
(x-y)(D^\nu_y)_{bc} \wt \sigma^{\mu\nu}_{ac} (y) - \wt
  \sigma^{\mu\nu}_{ab} (y) \partial^\nu_x \delta (x-y) .
\eeq
However, as a consequence of the gauge covariance of the covariant
current it holds that 
\beq
\frac{\delta \wt {\cal A}_b (y)}{\delta A^\mu_a (x)} \equiv
 (D^\nu_y)_{bc} {\cal F}^{\mu\nu}_{ac}(x,y)
\eeq
as may be checked explicitly \cite{Keln2}. Therefore, the coefficients in
(98) are not directly related to the anomaly and have to obey
\beq
\wt C^{\mu 0}_{ab}(y) = (D^\nu_y)_{bc}\wt \sigma^{\mu\nu}_{ac}(y)
\eeq
\beq
\wt S^{\mu 0k}_{ab}(y) = -\wt \sigma^{\mu k}_{ab}(y)
\eeq
\beq
\wt \sigma^{\mu 0}_{ab}(y) =0 
\eeq
and we find
\beq
\wt S^{00k}_{ab} (y) = \wt I^{k0}_{ab} (y)
\eeq
\beq
\wt C^{00}_{ab}(y) = \wt B^0_{ab}(y)
\eeq
and the consistency condition
\beq
\wt B^0_{ab}(y) = (D^\nu_y )_{bc} \wt I^{0\nu}_{ac}(y)
\eeq
which holds indeed, as may be checked by explicit computation
\cite{Keln2}. For the
anomalous part of the current commutator this leads to
\bdi
\delta (x^0 -y^0)[\wt J^0_a (x),\wt J^0_b (y)] -f_{abc}\wt J^0_c (y)
\delta (x-y) =
\edi
\beq
= \wt C^{00}_{ab}(y)\delta (x-y) +\wt S^{00k}_{ab}(y)\partial^k_x \delta
(x-y) \equiv \frac{\delta \wt {\cal A}_a (x)}{\delta A^0_b (y)} .
\eeq
Again, we have to calculate the $[X_b ,\wt J^0_a]$ commutators as in the
consistent case. However, the result is simply that each such term in
the Gauss operator commutator produces a contribution that is equal to
minus the above expression (106), see \cite{Keln2}. Therefore we find for
the covariant Gauss operator
\beq
\wt G_a (x) = \wt J^0_a (x) + {\bf X}_a (x)
\eeq
the Schwinger term
\beq
\wt {\cal G}_{ab}(x,y):= [\wt G_a (x),\wt G_b (y)] - f_{abc}\wt G_c (x)
{\vec \delta}(x-y) = -\int dy^0 \frac{\delta \wt {\cal A}_a (x)}{\delta 
A^0_b (y)}
\eeq
where the $dy^0$ integration just serves to get rid of the unwanted
$y^0$ dependence (this just kills a $\delta (x^0 -y^0)$, because there
is no time derivative in the above expression (108)).

\subsection{Explicit evaluation}

Now we are in a position to explicitly compute the Schwinger terms both
for $d=2$ and $d=4$. Starting with the $d=2$ case, we find from (58) and
(82) for the consistent ST 
\bdi
{\cal G}_{ab}(x,y) =-c_1 \epsilon^{0k} f_{bdc}A^k_d (y) \tr \lambda_a
\lambda_c  \delta^{(1)}(x-y)
\edi
\beq
=-c_1 \epsilon^{0k} \tr \lambda_a [\lambda_b , A^k ] \delta^{(1)}(x-y)
\eeq
($\delta^{(1)}(x-y) := \delta (x^1 -y^1)$) and for the covariant ST,
using (85) and (108)
\beq
\wt {\cal G}_{ab}(x,y)=-2c_1 \epsilon^{0k} \tr \lambda_a (\lambda_b
\partial^k_y \delta^{(1)} (x-y) + [\lambda_b , A^k ]\delta^{(1)}(x-y))
\eeq
where here and in the following functions always depend on $y$ when the
coordinate argument is not written down explicitly. In order to compare
with the expressions of Section 2, we omit $\epsilon^{0k}$ and multiply
by $(1/2)dy^k v_a (x) v_b (y)$, in the indicated order. Here $dy^k$ is a
one-form, $v_a (x)$ is a ghost, and all these objects {\em
  anti-commute}, e.g., $dy^k v_a (x) = -v_a (x) dy^k$. We find ($ A(y)
:= A^k (y) dy^k$, $v(x) := v_a (x) \lambda_a$)
\beq
{\cal G}(x,y)= -\frac{c_1}{2}\tr (v(x)v(y) + v(y)v(x))A(y)\delta^{(1)}
(x-y) 
\eeq
or, after integrating w.r.t. $\int dx^1 dy^1$
\beq
{\cal G} = -c_1 \int dy\, \tr v^2 A .
\eeq
In the same fashion, we get for $\wt {\cal G}_{ab}(x,y)$ ($d_y :=dy^k
\partial^k_y $)
\beq
\wt {\cal G}(x,y) = -c_1 \tr v(x)\Bigl( -(d_y \delta^{(1)} (x-y))v(y) +
\delta^{(1)} (x-y)(A(y)v(y) +v(y)A(y))\Bigr) 
\eeq
and (where a partial integration has to be performed)
\beq
\wt {\cal G} = -c_1 \int dy \, \tr v(dv +Av+vA) = -c_1 \int dy\, \tr vDv .
\eeq
Comparing with eq. (6) (for the consistent case) and Table 1 (for the
covariant case), we find that the relative sign of ${\cal G}$ and
$\wt {\cal G}$ is in precise agreement. 

For the case $d=4$ we find from (59) and (82) 
\bdi
{\cal G}_{ab}(x,y) = -\frac{c_2}{2}\epsilon^{0klm}\tr \Bigl( [\lambda_a
,\lambda_b ](\partial^k A^l A^m + A^k \partial^l A^m + A^k A^l A^m)
\edi
\beq
+(\lambda_b A^k \lambda_a - \lambda_a A^k \lambda_b )\partial^l A^m
\Bigr) \delta^{(3)}(x-y)
\eeq
(each derivative acts only on its immediate right hand neighbour), or,
after omitting $\epsilon^{0lkm}$ and multiplying by $(1/2)dy^k dy^l dy^m
v_a (x)v_b (y)$
\bdi
{\cal G}(x,y)= -\frac{c_2}{4}\tr \Bigl( (v(x)v(y) + v(y)v(x))(dAA + AdA
+A^3) 
\edi
\beq
+(v(y) A v(x) + v(x) A v(y))dA\Bigr) \delta^{(3)}(x-y)
\eeq
and upon integration $\int d^3 x d^3 y$
\beq
{\cal G}=-\frac{c_2}{2}\int dy\, \tr (v^2 (dAA+AdA +A^3)+vAvdA ).
\eeq
For the covariant ST we find from (86) and (108)
\bdi
\wt {\cal G}_{ab}(x,y) = -3c_2 \epsilon^{0klm}\tr \lambda_a \Bigl(
(\lambda_b \partial^k_y \delta^{(3)}(x-y) + (\lambda_b A^k - A^k
\lambda_b) \delta^{(3)}(x-y))(\partial^l A^m +A^l A^m)
\edi
\beq
+(\partial^l A^m +A^l A^m)(\lambda_b \partial^k_y \delta^{(3)}(x-y) +
(\lambda_b A^k - A^k \lambda_b )\delta^{(3)}(x-y))\Bigr)
\eeq
and
\bdi
\wt {\cal G}(x,y)= -\frac{3c_2}{2}\tr (-v(x))\Bigl( ((d_y
\delta^{(3)}(x-y)) v(y)-(vA + Av)\delta^{(3)}(x-y))(dA + A^2)
\edi
\beq
+(dA +A^2)((d_y \delta^{(3)}(x-y))v(y) -(vA+Av)\delta^{(3)}(x-y)) \Bigr)
\eeq
\bdi
\wt {\cal G}=-\frac{3c_2}{2}\int dy\, \tr v((dv + Av + vA)(dA +A^2)+(dA
+A^2)( dv + Av +vA))
\edi
\beq
=-\frac{3c_2}{2}\int dy\, \tr v(Dv F+FDv).
\eeq
Again, the relative sign of consistent and covariant ST precisely agrees
with the one in eq. (6) (consistent case) and Table 1 (covariant case).

\section*{Acknowledgement}

This work started at the Erwin Schr\"odinger Institute (ESI) 
program "Quantization, generalized BRS
cohomology and anomalies" and we would like to thank the ESI for kind
hospitality.
Further, we thank R.A.  Bertlmann for helpful discussions.
One of us (T.S.) is grateful to R.A. Bertlmann for the possibility to
spend some time in the creative atmosphere of the ESI.

\section*{Appendix}

\appendix

\section{The Schwinger terms of the Gauss Laws}
We start with the covariant case and we consider only the non-canonical
part of the commutator:
\begin{eqnarray}
\left[ \tilde{G}^a\left( x\right) ,\tilde{G}^b\left( y\right) \right]
_{n.c.} &=&\left[ X^a\left( x\right) +j^{0a}\left( x\right) ,X^b\left(
y\right) +j^{0b}\left( y\right) \right] _{n.c.}=  \nonumber \\
&=&\left[ X^a\left( x\right) ,X^b\left( y\right) \right] _{n.c.}+\left[
X^a\left( x\right) ,j^{0b}\left( y\right) \right] _{n.c.}+  \nonumber \\
&&+\left[ j^{0a}\left( x\right) ,X^b\left( y\right) \right]
_{n.c.}+\left[
j^{0a}\left( x\right) ,j^{0b}\left( y\right) \right] _{n.c.}.
\end{eqnarray}
The gauge field is an external field, therefore the commutator
\begin{equation}
\left[ X^a\left( x\right) ,X^b\left( y\right) \right] _{n.c.}
\end{equation}
is zero.
For the VEV of the commutator
\begin{equation}
\left[ j^{0a}\left( x\right) ,j^{0b}\left( y\right) \right] _{n.c.}
\end{equation}
we get, after some manipulations,
\begin{eqnarray}
\left\langle \left[ j^{0a}\left( x\right) ,j^{0b}\left( y\right)
\right] _{n.c.}\right\rangle  &=&{\mathrm tr}e^{-\left(
\varepsilon /2\right) \Delta _{y}}P_{-}\left( t,y,x\right)
t^{a}\left[ e^{-\left( \varepsilon /2\right) \Delta _{x}}-1\right]
\delta \left( x-y\right) t^{b}-  \nonumber \\ &&-\,\left(
x,a\leftrightarrow y,b\right), \label{SVEV of current commutator}
\end{eqnarray}
where $P_{-}\left( t,x,y\right) $ denotes the projection operator
(see Appendix C)
\begin{equation}
P_{-}\left( t,x,y\right) =\sum_{E_{n}<0}\zeta _{n}^{+}\left( t,x\right)
\zeta _{n}\left( t,x\right) .  \label{projection
operator}
\end{equation}
Then (\ref{SVEV of current commutator}) gives
\begin{eqnarray*}
&&{\mathrm tr}e^{-\varepsilon \Delta _{y}}P_{-}^{\left( 0\right)
}\left( y,x\right) t^{a}\left[ e^{-\varepsilon \Delta
_{x}}-1\right] \delta \left( x-y\right) t^{b}-\left(
x,a\leftrightarrow y,b\right) = \nonumber \\ && \nonumber \\
&=&\alpha\ {\mathrm tr}\int dE\theta \left( -E\right)
e^{-\varepsilon \Delta _{y}}e^{-iE(x-y)}t^{a}\times  \\ &&\qquad
\quad \times \int dq\left[ e^{-\varepsilon \Delta _{x}}-1\right]
e^{-iq(x-y)}t^{b}-\left( x,a\leftrightarrow y,b\right) \left.
=\right.  \\ &=&\alpha \,{\mathrm tr}\int dEdq\theta \left(
-E\right) e^{-\varepsilon \Delta _{y}}e^{-iE(x-y)}t^{a}\left[
e^{-\varepsilon \Delta _{x}}-1\right] e^{-iq(x-y)}t^{b}- \\
&&\qquad -\left( x,a\leftrightarrow y,b\right) \left. =\right.  \\
&=&\alpha \,{\mathrm tr}\int dEdq\theta \left( -E\right)
e^{-\varepsilon E^{2}}e^{-iE(x-y)}\left( 1-2i\varepsilon EA\left(
y\right) \right) t^{a}\times  \\ &&\qquad \quad \times \left[
\left( 1+2i\varepsilon qA\left( x\right) \right) e^{-\varepsilon
q^{2}}-1\right] e^{-iq(x-y)}t^{b}-\left( x,a\leftrightarrow
y,b\right) \left. =\right.
\end{eqnarray*}
\begin{eqnarray}
&=&\alpha \,\int dEdq\theta \left( -E\right) e^{-\varepsilon
E^{2}}e^{-i\left( E+q\right) (x-y)}\left[ e^{-\varepsilon
q^{2}}-1\right]
\cdot {\mathrm tr}t^{a}t^{b}-\left( x,a\leftrightarrow y,b\right) -
\nonumber
\\
&&-\ i\alpha \,\int dEdq\theta \left( -E\right) e^{-\varepsilon
  E^{2}}\left[
e^{-\varepsilon q^{2}}-1\right] e^{-i\left( E+q\right)
  (x-y)}2\varepsilon
E\cdot {\mathrm tr}A\left( y\right) t^{a}t^{b}+  \nonumber \\
&&+\left( x,a\leftrightarrow y,b\right) +  \nonumber \\
&&+\ i\alpha \,\int dEdq\theta \left( -E\right) e^{-\varepsilon
E^{2}}2\varepsilon qe^{-\varepsilon q^{2}}\cdot {\mathrm tr}t^{a}A\left(
x\right) t^{b}-\left( x,a\leftrightarrow y,b\right) \left. =\right.
\nonumber \\
&&  \nonumber \\
&=&\alpha \,\int d\xi e^{-i\xi (x-y)}\int dE\theta \left( -E\right)
e^{-\varepsilon E^{2}}\left[ e^{-\varepsilon \left( \xi -E\right)
  ^{2}}-1%
\right] \cdot {\mathrm tr}t^{a}t^{b}-\left( x,a\leftrightarrow y,b\right)
-
\nonumber \\
&&-\ i\alpha \,\int d\xi e^{-i\xi (x-y)}\int dE\theta \left( -E\right)
2\varepsilon Ee^{-\varepsilon E^{2}}\left[ e^{-\varepsilon \left( \xi
-E\right) ^{2}}-1\right] \cdot {\mathrm tr}A\left( y\right) t^{a}t^{b}+
\nonumber \\
&&+\left( x,a\leftrightarrow y,b\right) +  \nonumber \\
&&+\ i\alpha \,\int d\xi e^{-i\xi (x-y)}\int dE\theta \left( -E\right)
2\varepsilon qe^{-\varepsilon E^{2}}e^{-\varepsilon \left( \xi -E\right)
^{2}}\cdot {\mathrm tr}t^{a}A\left( x\right) t^{b}-  \nonumber \\
&&-\left( x,a\leftrightarrow y,b\right) \left. =\right.   \nonumber \\
&&  \nonumber \\
&=&\alpha \,\int d\xi e^{-i\xi (x-y)}\int dEe^{-\varepsilon E^{2}}\left[
e^{-\varepsilon \left( \xi -E\right) ^{2}}-1\right] \left( \theta \left(
-E\right) -\theta \left( E\right) \right) \cdot {\mathrm tr}t^{a}t^{b}-
\nonumber \\
&&-\ i\alpha \,\left\{ \left[ \int d\xi e^{-i\xi (x-y)}\int dE\theta
\left(
-E\right) 2\varepsilon Ee^{-\varepsilon E^{2}}\left[ e^{-\varepsilon
  \left(
\xi -E\right) ^{2}}-1\right] - \right. \right.   \nonumber \\
&&-\left. \int d\xi e^{-i\xi (x-y)}\int dE\theta \left( E\right)
2\varepsilon \left( \xi -E\right) e^{-\varepsilon E^{2}}e^{-\varepsilon
\left( \xi -E\right) ^{2}}\right] \cdot {\mathrm tr}A\left( y\right)
t^{a}t^{b} -  \nonumber \\
&&
\Bigl. -\left( x,a\leftrightarrow y,b\right) \Bigr \}
\,\left. \stackrel{\varepsilon \rightarrow 0}{\sim }\right.
\nonumber
\\
&&  \nonumber \\
&\sim &-\ \alpha \,\int d\xi e^{-i\xi (x-y)}\int dEe^{-\varepsilon
  E^{2}}
\left[ e^{-\varepsilon E^{2}}\left( 1-\varepsilon \xi ^{2}+2\varepsilon
  E\xi
\right) -1\right] \varepsilon (E)\cdot {\mathrm tr}t^{a}t^{b}-  \nonumber
\\
&&-\ i\alpha \,\left\{ \int d\xi e^{-i\xi (x-y)}\times \right.
\nonumber \\
&&
\begin{array}{lllll}
\qquad \times \left[ \int dE2\varepsilon Ee^{-\varepsilon
E^{2}}e^{-\varepsilon \left( \xi -E\right) ^{2}}\left( \theta \left(
-E\right) +\theta \left( E\right) \right) - \right.  &  &  &  &
\backslash \,%
\stackrel{\varepsilon \rightarrow 0}{\rightarrow }0 \\
\qquad \quad -\int dE\theta \left( -E\right) 2\varepsilon
Ee^{-\varepsilon
E^{2}}- &  &  &  & \backslash \,\stackrel{\varepsilon \rightarrow 0}{%
\rightarrow }1 \\
\qquad \quad \left. -\int dE\theta \left( E\right) 2\varepsilon \xi
e^{-\varepsilon E^{2}}e^{-\varepsilon \left( \xi -E\right) ^{2}}\right]
\times  &  &  &  & \backslash \,\stackrel{\varepsilon \rightarrow 0}{%
\rightarrow }0
\end{array}
\nonumber \\
&&\qquad \times {\mathrm tr}A\left( y\right) t^{a}t^{b}\,-\left(
x,a\leftrightarrow y,b\right)
\Bigl. \Bigr \}%
%
\left. \sim \right.   \nonumber \\ &&  \nonumber \\ &\sim &-\
\alpha \,\int d\xi \xi e^{-i\xi (x-y)}\int dEe^{-2\varepsilon
E^{2}}2\varepsilon E\varepsilon \left( E\right) \cdot
{\mathrm tr}t^{a}t^{b}- \nonumber \\ &&-\ i\alpha \,\left\{ \int
d\xi e^{-i\xi (x-y)}\cdot {\mathrm tr}A\left( y\right)
t^{a}t^{b}-\int d\xi e^{-i\xi (x-y)}\cdot {\mathrm tr}A\left(
x\right) t^{b}t^{a}\right\} \left. \stackrel{\varepsilon \rightarrow
  0}{=}%
\right.   \nonumber \\
&&  \nonumber \\
&=&-\ \alpha \,\int d\xi e^{-i\xi (x-y)}\xi \cdot
{\mathrm tr}t^{a}t^{b}+2\pi
i\alpha \,\delta \left( x-y\right) \cdot {\mathrm tr}t^{a}\left[ A\left(
y\right) ,t^{b}\right] =  \nonumber \\
&&  \nonumber \\
&=&-\frac{i}{2\pi }\,\partial _{x}\delta (x-y)\cdot
{\mathrm tr}t^{a}t^{b}+%
\frac{i}{2\pi }\,\delta (x-y)\cdot {\mathrm tr}t^{a}\left[ A\left(
y\right) ,t^{b}\right],
\end{eqnarray}
where $\alpha = 1/\left( 2\pi \right)^{2}$. Then
\begin{eqnarray}
\left\langle \left[ j^{0a}\left( x\right) ,j^{0b}\left( y\right) \right]
_{n.c.}\right\rangle  &=&-\frac{i}{2\pi }\partial _{x}\delta
(x-y)\cdot
{\mathrm tr}t^{a}t^{b}+\frac{i}{2\pi }\delta (x-y)\cdot
{\mathrm tr}t^{a}\left[
A\left( y\right) ,t^{b}\right]   \nonumber \\
&&
\end{eqnarray}
and the covariant Schwinger term of the commutator of the full
Gauss law operators has the form
\begin{eqnarray}
\widetilde{ST}^{ab} &=&\left\langle \left[ \tilde{G}^{a}\left( x\right)
,
\tilde{G}^{b}\left( y\right) \right] _{n.c.}\right\rangle
=-\left\langle
\left[ j^{0a}\left( x\right) ,j^{0b}\left( y\right) \right]
_{n.c.}\right\rangle =  \nonumber \\
&=&\frac{i}{2\pi }\partial _{x}\delta (x-y)\cdot
{\mathrm tr}t^{a}t^{b}+\frac{%
i}{2\pi }\delta (x-y)\cdot {\mathrm tr}t^{a}\left[ A_{1}\left( y\right)
,t^{b}%
\right] ,
\end{eqnarray}
where we used the result for the cross-term
\begin{equation}
\left\langle \left[ X^{a}\left( x\right) ,j^{0b}\left( y\right) \right]
\right\rangle =\frac{i}{2\pi }\left( \partial _{1}^{x}\delta
^{ac}+f^{aec}A_{1}^{e}\left( x\right) \right) \delta \left( x-y\right)
\cdot
{\mathrm tr}t^{b}t^{c}\,
\end{equation}
obtained in Appendix B.

For the commutator of the consistent Gauss laws we get
\begin{eqnarray}
&&\left[ G^{a}\left( x\right) ,G^{b}\left( y\right) \right] =  \nonumber
\\
&=&\left[ \tilde{G}^{a}\left( x\right) ,\tilde{G}^{b}\left( y\right)
\right]
+\left[ \tilde{G}^{a}\left( x\right) ,\Delta j^{0b}\left( y\right)
\right] +%
\left[ \Delta j^{0a}\left( x\right) ,\tilde{G}^{b}\left( y\right)
\right] =
\nonumber \\
&=&f^{abc}\tilde{G}^{c}\left( x\right) \delta \left( x-y\right) =
\nonumber
\\
&=&f^{abc}G^{c}\left( x\right) \delta \left( x-y\right) -f^{abc}\Delta
j^{0c}\left( y\right) \delta \left( x-y\right) ,
\end{eqnarray}
where we used the equality
\begin{equation}
\widetilde{ST}^{ab}+\left[ \tilde{G}^{a}\left( x\right) ,\Delta
j^{0b}\left(
y\right) \right] +\left[ \Delta j^{0a}\left( x\right)
,\tilde{G}^{b}\left(
y\right) \right] =0.
\end{equation}
which results from
\begin{eqnarray}
\left[ \tilde{G}^{a}\left( x\right) ,\Delta j^{0b}\left( y\right)
\right] &=&\left[ X^{a}\left( x\right) ,\Delta j^{0b}\left(
y\right) \right] =X^{a}\left( x\right) \Delta j^{0b}\left(
y\right) =  \nonumber \\ &=&\frac{i}{4\pi }\varepsilon ^{01}\left(
\delta ^{ac}\partial ^{\mu }+f^{aec}A^{\mu e}\left( x\right)
\right) \frac{\delta }{\delta A^{\mu c}\left( x\right)
}{\mathrm tr}\left( t^{b}A_{1}\left( y\right) \right) = \nonumber
\\ &=&\frac{i}{4\pi }\left( \delta ^{ac}\partial
_{x}^{1}+f^{aec}A^{1e}\left( x\right) \right)\delta \left(
x-y\right) \cdot {\mathrm tr}t^{b}t^{c} .
\end{eqnarray}
Therefore
\begin{eqnarray}
ST^{ab} &=&-f^{abc}\Delta j^{0c}\left( y\right) \delta \left( x-y\right)
=
\nonumber \\
&=&\frac{i}{4\pi }\varepsilon ^{0\nu }{\mathrm tr}\left(
f^{abc}t^{c}A_{\nu
}\right) \delta \left( x-y\right) =  \nonumber \\
&=&\frac{i}{4\pi }\delta \left( x-y\right) \cdot {\mathrm tr}\left(
\left[
t^{a},A_{1}\right] t^{b}\right) .
\end{eqnarray}

\section{The cross-term}

For the VEV of the cross-term
\begin{equation}
\left[ X^{a}\left( x\right) ,j^{0b}\left( y\right) \right]
\end{equation}
we get
\begin{eqnarray}
\left\langle \left[ X^{a}\left( x\right) ,j^{0b}\left( y\right) \right]
\right\rangle  &=&-\left\langle \left( \partial _{x}^{1}\delta
^{ac}+f^{aec}A^{1e}\left( x\right) \right) \frac{\delta }{\delta
A^{1c}\left( x\right) }j^{0b}\left( y\right) \right\rangle =
\nonumber
\\
&=&-\left( \partial _{x}^{1}\delta ^{ac}+f^{aec}A^{1e}\left( x\right)
\right) \left\langle \frac{\delta }{\delta A^{1c}\left( x\right) }%
j^{0b}\left( y\right) \right\rangle .
\end{eqnarray}
Because
\begin{eqnarray}
&&\left( \frac{\delta }{\delta A^{1c}\left( x\right) }e^{-\left(
  \varepsilon
/2\right) \Delta _{y}}\right) P_{-}(y,z)\left. 
\stackrel{\varepsilon \rightarrow 0}{
\sim } \right.
\nonumber \\
&\sim &\left( \frac{\delta }{\delta A^{1c}\left( x\right) }e^{-\left(
\varepsilon /2\right) \Delta _{y}}\right) P_{-}^{\left( 0\right) }\left(
y,z\right) =  \nonumber \\
&=&\frac{1}{2\pi }\left( \frac{\delta }{\delta A^{1c}\left( x\right) }%
e^{-\left( \varepsilon /2\right) \Delta _{y}}\right) \int dE\theta
\left(
-E\right) e^{iE\left( y-z\right) }
 \stackrel{\varepsilon \rightarrow 0}{
= } \nonumber \\
&=&-\frac{i}{2\pi }\int dE\theta \left( -E\right) \varepsilon
Ee^{iE\left(
y-z\right) }e^{-\left( \varepsilon /2\right) p^{2}}\delta \left(
x-y\right)
t^{c}
\end{eqnarray}
and
\begin{eqnarray}
&&-\frac{i}{2\pi }\int dE\theta \left( -E\right) \varepsilon
Ee^{iE\left( y-z\right) }e^{-\left( \varepsilon /2\right)
E^{2}}e^{-\left( \varepsilon /2\right) \stackrel{\leftarrow
}{\Delta}_{z}} \nonumber \\ &=&-\frac{i}{2\pi }\int dE\theta
\left( -E\right) \varepsilon Ee^{iE\left( y-z\right)
}e^{-\varepsilon E^{2}},
\end{eqnarray}
we obtain
\begin{eqnarray}
&&\left\langle \frac{\delta }{\delta A^{1c}\left( x\right)
}j^{0b}\left( y\right) \right\rangle \left. =\right. \nonumber
\\ &=&\lim_{z\rightarrow y}{\mathrm tr}t^{b}\left[ \left(
\frac{\delta }{\delta A^{1c}\left( x\right) }e^{-\left(
\varepsilon /2\right) \Delta _{y}}\right) P_{-}(y,z)e^{-\left(
\varepsilon /2\right) \stackrel{\leftarrow }{\Delta}_{z}}\right. +
\nonumber \\ &&+\left. e^{-\left( \varepsilon /2\right) \Delta
_{y}}P_{-}(y,z)\left( \frac{\delta }{\delta A^{1c}\left( x\right)
}e^{-\left( \varepsilon /2\right) \stackrel{\leftarrow
}{\Delta}_{z}}\right) \right] =  \nonumber \label{kuk} \\
&=&-\frac{i}{\pi }\int \theta \left( -E\right) \varepsilon
Ee^{-\varepsilon E^{2}}dE\,\delta \left( x-y\right) \cdot
{\mathrm tr}t^{b}t^{c}.
\end{eqnarray}
The integral is
\begin{equation}
\int \theta \left( -E\right) \varepsilon Ee^{-\varepsilon
  E^{2}}dE=-\frac{1}{%
2}
\end{equation}
and therefore
\begin{equation}
\left\langle \frac{\delta }{\delta A^{1c}\left( x\right) }j^{0b}\left(
y\right) \right\rangle =\frac{i}{2\pi }\,\delta \left( x-y\right)
\cdot
{\mathrm tr}t^{b}t^{c}.
\end{equation}
So, for the commutator
\[
\left\langle \left[ X^{a}\left( x\right) ,j^{0b}\left( y\right) \right]
\right\rangle 
\]
we finally get
\begin{equation}
\left\langle \left[ X^{a}\left( x\right) ,j^{0b}\left( y\right) \right]
\right\rangle =\frac{i}{2\pi }\left( \partial _{1}^{x}\delta
^{ac}+f^{aec}A_{1}^{e}\left( x\right) \right) \,\delta \left( x-y\right)
\cdot {\mathrm tr}t^{b}t^{c}.
\end{equation}

\section{The projection operator}

For our purposes we expand the projection operator (\ref{projection
  operator})
\begin{eqnarray}
P_{-}\left( t,x,y\right) &=&\left\langle x\right|
\oint_{C_{-}}\frac{dE}{%
2\pi i}\frac 1{E-H\left( t\right) }\left| y\right\rangle =  \nonumber \\
&=&\left\langle x\right| \oint_{C_{-}}\frac{dE}{2\pi i}\frac
1{E-H_0-V\left(
t\right) }\left| y\right\rangle =  \nonumber \\
&=&\left\langle x\right| \oint_{C_{-}}\frac{dE}{2\pi i}\frac
1{E-H_0}\left|
y\right\rangle +  \nonumber \\
&&+\left\langle x\right| \oint_{C_{-}}\frac{dE}{2\pi i}\frac
1{E-H_0}V\left(
t\right) \frac 1{E-H_0}\left| y\right\rangle +\ldots \left. =\right.
\nonumber \\
&=&P_{-}^{\left( 0\right) }\left( x,y\right) +P_{-}^{\left( 1\right)
  }\left(
t,x,y\right) +\ldots ,  \label{series of projection operator}
\end{eqnarray}
where $C_{-}$ is a contour surrounding the negative real axis in the
complex
$E$ plane.

For the calculation of the commutators it is sufficient to consider only
the
first term of (\ref{series of projection operator})
\begin{eqnarray}
P_{-}^{\left( 0\right) }\left( x,y\right) &=&\sum_{E_n<0}\zeta _n\left(
x\right) \zeta _n^{+}\left( y\right) = \\
&=&\frac 1{2\pi }\int dE\theta \left( -E\right) e^{iE(x-y)}.
\end{eqnarray}

\end{document}